%% file: 0_main.tex
\documentclass[10pt,conference]{IEEEtran}
\usepackage{cite}
\usepackage{amsmath,amssymb,amsfonts}
\usepackage{algorithmic}
\usepackage{graphicx}
\usepackage{textcomp}
\usepackage{xcolor}
\usepackage{listingsutf8}
\usepackage{url}
\usepackage{hyperref}
\usepackage[capitalise, noabbrev]{cleveref}
\usepackage{booktabs}
\usepackage{balance}
\usepackage[all]{nowidow}

\usepackage{tikz}
\usetikzlibrary{shapes,backgrounds,patterns}

\usepackage{lipsum}

\input{commands}

\def\BibTeX{{\rm B\kern-.05em{\sc i\kern-.025em b}\kern-.08em
    T\kern-.1667em\lower.7ex\hbox{E}\kern-.125emX}}
\begin{document}

\title{BeFaaS: An Application-Centric Benchmarking Framework for FaaS Platforms}


\author{\IEEEauthorblockN{Martin Grambow, Tobias Pfandzelter, Luk Burchard, Carsten Schubert, Max Zhao, David Bermbach}
    \IEEEauthorblockA{\textit{TU Berlin \& Einstein Center Digital Future} \\
        \textit{Mobile Cloud Computing Research Group}\\
        Berlin, Germany \\
        \{mg, tp, lubu, casc, mazh, db\}@mcc.tu-berlin.de}
}

\maketitle
\begin{abstract}
Following the increasing interest and adoption of FaaS systems, benchmarking frameworks for determining non-functional properties have also emerged. 
While existing (microbenchmark) frameworks only evaluate single aspects of FaaS platforms, a more holistic, application-driven approach is still missing. 

In this paper, we design and present BeFaaS, an extensible application-centric benchmarking framework for FaaS environments that focuses on the evaluation of FaaS platforms through realistic and typical examples of FaaS applications.
BeFaaS includes a built-in e-commerce benchmark, is extensible for new workload profiles and new platforms, supports federated benchmark runs in which the benchmark application is distributed over multiple providers, and supports a fine-grained result analysis. 

Our evaluation compares three major FaaS providers in single cloud provider setups and shows that BeFaaS is capable of running each benchmark automatically with minimal configuration effort and providing detailed insights for each interaction.
\end{abstract}


\input{1_intro}
\input{2_related}
\input{3_requirements}
\input{4_design}
\input{5_impl}
\input{6_evaluation}
\input{7_discussion}
\balance
\input{8_conclusion}

\section*{Acknowledgments}
The authors would like to thank Emily Dietrich and Christoph Witzko who also contributed to the implementation of the BeFaaS prototype within the scope of a master’s project.

\bibliographystyle{IEEEtran}
\bibliography{bibliography}
\end{document}

%% file: commands.tex
\newcommand\free[0]{\vspace{1em}\noindent}

%% file: 1_intro.tex
\section{Introduction}

All major cloud providers offer Function-as-a-Service (FaaS) solutions where users only have to take care of their source code (functions) while the underlying infrastructure and environment is abstracted away by the provider. 
FaaS-based applications are split by their business functionality into individual functions which are deployed on a FaaS platform which, e.g., handles the execution and automatic scaling.
The developer does not have any direct control over the infrastructure and can only define high-level parameters, such as the region in which the function should run. 
This complicates an already challenging comparison of cloud providers~\cite{paper_leitner_cloud_variability,paper_bermbach_expect_the_unexpected}, as the cloud variability is further compounded by an additional, unknown infrastructure component. 

Existing work dealing with benchmarking of FaaS platforms focuses on the execution of small, so-called microbenchmarks which deploy and call a simple function (e.g., a matrix multiplication~\cite{back_using_2018} or a random number generator~\cite{malawski_benchmarking_2017}).
While microbenchmarks are useful for studying and comparing specific characteristics, they can give only focused and limited insights into the platform behavior that applications can expect~\cite{book_cloud_service_benchmarking}.
An application-centric benchmark, in contrast, mimics the behavior of a realistic application while closely observing the platform behavior.
This allows developers to better compare different service options, a strategy also taken by the TPC benchmarks\footnote{\url{www.tpc.org}}.
To the best of our knowledge, such an application-centric benchmark for FaaS platforms does not exist yet.

To address this gap, we here propose \emph{BeFaaS}, an extensible framework for executing application-centric benchmarks against FaaS platforms that includes a realistic e-commerce example benchmark.
BeFaaS is also the first benchmarking framework with out-of-the-box support for federated cloud~\cite{paper_kurze_cloud_federation} setups which allows us to evaluate complex configurations in which an application is distributed over multiple FaaS platforms running on a mixture of cloud, edge, and fog nodes.
Beyond this, BeFaaS is focused on ease-of-use and collects fine-grained measurements which can be used for a detailed post-experiment drill-down analysis, e.g., to identify cold starts or other request-level effects.

In this regard, we make the following contributions:
\begin{itemize}
    \item We derive requirements for an application-centric FaaS benchmarking framework.
    \item We propose BeFaaS, an extensible framework for the execution of application-centric FaaS benchmarks and describe our example benchmark.
		\item We present our proof-of-concept prototype which is available as open source and currently supports six FaaS platforms.
    \item We run a number of experiments and use them to compare three public FaaS offerings.
\end{itemize}

This paper is structured as follows: After outlining the related work in \cref{sec:rw} and deriving the requirements for an application-centric FaaS benchmark in \cref{sec:requirements}, we present the design, architecture, and features of BeFaaS in \cref{sec:design}. 
Next, we describe our implementation of BeFaaS including the built-in e-commerce benchmark in \cref{sec:impl} which we then use to evaluate three FaaS platforms (\cref{sec:evaluation}).
Finally, we discuss the current limitations and future work in \cref{sec:discussion} before concluding in \cref{sec:conclusion}.

%% file: 2_related.tex
\section{Related Work}
\label{sec:rw}

Existing research on benchmarking of FaaS environments has so far focused on microbenchmarks. 
Application-centric benchmarks that consider the overall performance of multiple functions, the interaction with external services, and the effects of different application load profiles are mostly still missing.

Microbenchmarks call single functions repeatedly and evaluate the resulting metrics. 
These functions are often designed for a specific purpose, e.g., to stress the CPU of the test system or to evaluate the test system with a disk-intensive workload. 
Multiple performance evaluation studies are based on microbenchmarks which compare FaaS vendors, e.g.,~\cite{wang_peeking_2018, malawski_benchmarking_2017, figiela_performance_2018, manner_cold_2018, lee_evaluation_2018, back_using_2018, yu_characterizing_2020, martins_benchmarking_2020}. 
Besides scaling of functions, cold start latency, and instance lifetimes, the studies also evaluate metrics such as CPU utilization, network throughput, and costs. 
Almost all experiments, however, focus on single isolated aspects and do not create comparability of platforms for FaaS application developers.

Some studies also consider more complex applications such as image processing~\cite{kim_functionbench_2019}, analyze chained functions, or deploy real world applications on serverless platforms~\cite{yu_characterizing_2020}.
While these papers also use application-centric workloads for experiments, their goal was not to propose a comprehensive framework for the execution of application-centric FaaS benchmarks.

PanOpticon~\cite{somu_panopticon_2020} uses a deployment, workload, and metrics module to evaluate chained functions and a simple chat server on two different FaaS vendors. 
Although PanOpticon has similar goals as BeFaaS, it neither supports detailed drill-down analysis nor federated multi-provider setups.
Also, van Eyk et al.~\cite{van_eyk_beyond_2020} developed a high-level architecture and stated requirements for serverless benchmarking.
While their project has a similar goal as BeFaaS, it unfortunately seems to still be in a vision state.
Existing preliminary source code components are, according to the paper, not available online, whereas we publish BeFaaS as an open-source research prototype.

Beyond FaaS, there are a number of application-centric benchmarking frameworks in other domains, e.g., for database and storage systems~\cite{bermbach_benchfoundry_2017,difallah2013oltp} or for virtual machines~\cite{borhani2014wpress}. These can, however, not easily be adapted to FaaS platforms.

%% file: 3_requirements.tex
\section{Requirements}
\label{sec:requirements}
While microbenchmarks are highly useful for studying individual features of a system-under-test (SUT), application-centric benchmarks support end-to-end comparison of different platforms and configurations.
Aside from standard benchmarking requirements such as portability or fairness~\cite{huppler_art_2009, paper_bermbach_benchmarking_middleware,bermbach_benchfoundry_2017, folkerts_benchmarking_2013,book_cloud_service_benchmarking}, an application-centric FaaS benchmarking framework needs to fulfill a number of specific requirements which we describe in this section.

\free
\textbf{R1 -- Realistic Benchmark Application:}
The performance of a FaaS platform depends on the application that is deployed on it.
For instance, an application that frequently causes cold starts through a growing request rate will be better off on AWS Lambda while an application that frequently causes cold starts through short temporary load spikes will be better off on Apache OpenWhisk due to their different request queuing mechanisms~\cite{paper_bermbach_faas_coldstarts}.
This means that the benchmark application should be as close as possible to the real application for which the analysis is made~\cite{book_cloud_service_benchmarking}, e.g., in line with the findings of~\cite{shahrad_serverless_2020}.
A key requirement is, hence, that \emph{a FaaS benchmark should mimic real applications as closely as possible}.

\free
\textbf{R2 -- Extensibility for New Workloads:}
FaaS platforms are highly flexible and can be used for a wide variety of applications, so the world of FaaS applications is evolving rapidly.
As such, any set of ``typical'' FaaS applications -- and thus the workload profile for a FaaS platform -- can only be considered a snapshot in time.
Likewise, the load profiles of existing FaaS applications, i.e., the amount and type of requests that the application handles, are likely to evolve over time.
Therefore, we argue that \emph{a FaaS benchmarking framework should be easily extensible in terms of adding new benchmark applications and updating load profiles for existing benchmarks}.

\free
\textbf{R3 -- Support for Modern Deployments:}
FaaS is often used as the ``glue'' between cloud services, web APIs, and legacy systems.
Thus, a benchmarking framework must also consider these links and support external services. 
Furthermore, today's applications are often distributed over cloud, edge, and fog resources~\cite{paper_bermbach_fog_vision,paper_zhang_gdp, paper_pfandzelter_LEO_serverless}.
Here, for example, hybrid clouds can keep sensitive functions on premises while non-critical functions are hosted in a public cloud; similar setups exist for edge and fog computing use cases~\cite{paper_pallas_fog4privacy,grambow_public_2018, aslanpour_serverless_nodate}.
As such, assuming a single-cloud deployment is unrealistic for benchmarks aiming to be as similar as possible to realistic applications.
\emph{A benchmarking framework needs to support external services and federated setups in which application functions are deployed on one or more FaaS platforms distributed across cloud, edge, and fog}.

\free
\textbf{R4 -- Extensibility for New Platforms:}
Today, all major cloud service providers offer FaaS platforms and there is a growing range of open-source FaaS systems, for example, systems that specifically target the edge~\cite{paper_george_nanolambda,pfandzelter_tinyfaas_2020}.
As interfaces are constantly evolving and new platforms are introduced, a cross-platform benchmarking framework \emph{needs to be extensible to support future FaaS platforms}.

\free
\textbf{R5 -- Support for Drill-down Analysis:}
An application-centric FaaS benchmark can help to evaluate the suitability of different sets and configurations of FaaS platforms for a specific application.
What it can usually not provide are explanations for its finding, e.g., the different cold start management behavior of AWS Lambda and Apache OpenWhisk mentioned above~\cite{paper_bermbach_faas_coldstarts}.
To facilitate root cause analysis and help evaluators explain the patterns they see in the benchmark results, we argue that \emph{an application-centric FaaS benchmarking framework should support drill-down analysis by logging fine-grained measurement results including typical metrics of microbenchmarks}.

\free
\textbf{R6 -- Minimum Required Configuration Overhead:}
An application-centric FaaS benchmarking framework should be easy to use and provide reproducible results.
This includes configuration, deployment, execution, as well as collection and analysis of results, e.g., based on infrastructure automation.
Hence, \emph{a FaaS benchmarking framework should be designed to require as little manual effort as possible}.

%% file: 4_design.tex
\section{Design}
\label{sec:design}
In this section, we give an overview of the BeFaaS design, starting with an overview of the BeFaaS architecture and components (\cref{subsec:architecture}) before describing the key features of BeFaaS (\crefrange{subsec:design1}{subsec:design-last}).

\input{fig1_architecture.tex}

\subsection{Architecture and Components\label{subsec:architecture}}
In BeFaaS, the execution of functions of a benchmark application is the \emph{workload} that actually benchmarks the FaaS platform, i.e., executing a function creates \emph{stress} on the SUT.
Since functions do not ``self-start'' executing, we need an additional load generator that invokes the FaaS functions of our benchmark application; see also \cref{fig:arch} for a high-level architecture overview.

For a benchmark run, BeFaaS requires three inputs: (i) the source code of the FaaS functions forming the benchmark application, (ii) a load profile for the load generator, and (iii) a deployment configuration that describes the environment configuration for each function and FaaS platform (the SUTs).

For a benchmark run, application code and deployment configuration are initially converted into deployment artifacts by the \emph{Deployment Compiler}.
The Deployment Compiler instruments and wraps each function's code with BeFaaS library calls and injects vendor-specific instructions defined in deployment adapters which enables request tracing and fine-grained metrics.
The resulting deployment artifacts are passed to the \emph{Benchmark Manager}.

The Benchmark Manager orchestrates the experiment:
First, it sets up the \emph{SUT} by deploying each function based on the information in the respective artifact.
If there are external services, these can either be deployed by the Benchmark Manager as well or linked to the SUT using environment variables. 
In the second step, it initializes the \emph{Load Generator} with the workload information described in a load profile.
Then, the benchmark run is triggered and the Load Generator invokes the functions of the benchmark application which log every request in detail including timestamps, origin function, and called functions (if applicable).
Finally, once the benchmark run is completed, the Benchmark Manager collects the log files from all FaaS platforms used, aggregates them into a joint results file, and destroys all provisioned resources; see \cref{fig:overview} for an overview of the components in the BeFaaS framework and their interactions.

\input{fig2_overview.tex}

\subsection{Realistic Benchmarks\label{subsec:design1}}
To provide a relevant and realistic application-centric benchmark (\textbf{R1}), BeFaaS already comes with one built-in benchmark which mimics an e\nobreakdash-commerce scenario and represents a typical use case for FaaS applications (this application is explained in further detail in \cref{sec:impl}).
Our benchmark adheres to the empirical findings of Shahrad et al.~\cite{shahrad_serverless_2020}, is composed of several functions that interact with each other to form function chains, and uses external services such as a database system for persistence.
The benchmark application comes with a default load profile that covers all relevant aspects as well as several further load profiles to emphasize selected stress situations, e.g., to provoke more cold starts.
In combination, the benchmark represents a complete FaaS application: load balancing at the provider endpoint(s), interconnected calls of several functions, calls to external services such as database systems, and multiple load profiles which, e.g., provoke scaling of resources.

The modular design of BeFaaS, however, also allows us to easily add further benchmark applications and load profiles or to adapt existing ones to the concrete needs of the developer (\textbf{R2}).
For adding a new benchmark, the respective application only needs to use the BeFaaS library (described in \cref{sec:impl}) for function calls and to have unique function names.

\subsection{Benchmark Portability and Federated FaaS Deployments}
To support portability of benchmarks and federated deployments, BeFaaS relies on unique function names, individual deployment artifacts for every function, and a single endpoint for every deployed function (\textbf{R3}):
With globally unique function names, the endpoints of the deployed functions are already known during the compilation phase.
The Deployment Compiler maps these endpoints to the canonical function names (defined in the application) and compiles them into the source code.
Moreover, the compiler also injects endpoints to external services such as database systems using environment variables which were set in the respective setup script or defined manually.
This decouples the ability of a function to call another function or a platform service from its deployment location and enables BeFaaS to support arbitrarily complex deployments: it is indeed possible to run every function on a different FaaS platform.

Each FaaS platform offers a different interface for life-cycle and configuration management of functions.
As the smallest common interface, BeFaaS requires that each platform provides API-based access to (i) deploying functions, (ii) retrieving log entries from the standard logging interface, and (iii) removing functions.
The Deployment Compiler wraps this functionality using an adapter mechanism and selects the appropriate instructions for the target platform specified in the deployment configuration.
Additional FaaS platforms that fulfill this minimal interface can easily be added by implementing a corresponding adapter (\textbf{R4}).

\subsection{Detailed Request Tracing}
To enable a detailed drill-down analysis of experiment results (\textbf{R5}), the Deployment Compiler injects and wraps code that collects detailed measurements during the benchmark run:
The compiler adds timestamping to determine start, end, and latency of calls to functions and external services. 

Besides these timestamps, the compiler also injects code that generates context IDs and pair IDs to assign individual calls to their respective context later on.
Here, a context ID is generated once for each function chain (the first function call) which is propagated to every subsequent call to other functions.
To link the individual calls of a function chain, the compiler injects source code to create pair IDs of randomly generated keys that link calling and called function.
Thus, it is possible to trace every single request through the benchmark application and to generate call trees for every context and function chain.

Finally, to independently and reliably detect cold starts, the Deployment Compiler also injects code that evaluates a local environment variable on the executor at the provider side.
If this variable is not present, the function runs on a new executor (cold start), the variable is created, filled with a randomly generated key, and the cold start is logged.

All data that enable fine-grained results (timestamps, context IDs, pair IDs, and executor keys) are recorded on the console using the standard logging interface of the respective FaaS vendor.
Initial experiments with Amazon Web Services (AWS), Google Cloud Platform (GCP), and Microsoft Azure (Azure) have shown that the cost of logging is at most in the microsecond range.

\subsection{Automated Experiment Orchestration\label{subsec:design-last}}
The BeFaaS framework requires only the application code, a deployment configuration, and a load profile to automatically perform the benchmark experiment (\textbf{R6}).
First, all business logic, dependencies, and BeFaaS instrumentation logic are bundled into a single deployment artifact by the Deployment Compiler.
Next, the Benchmark Manager orchestrates the experiment and provides a simple interface for starting the benchmark run, monitoring its process, and collecting fine-grained results for further analysis.

%% file: fig1_architecture.tex
\begin{figure}
    \centering
    \includegraphics[width=\columnwidth]{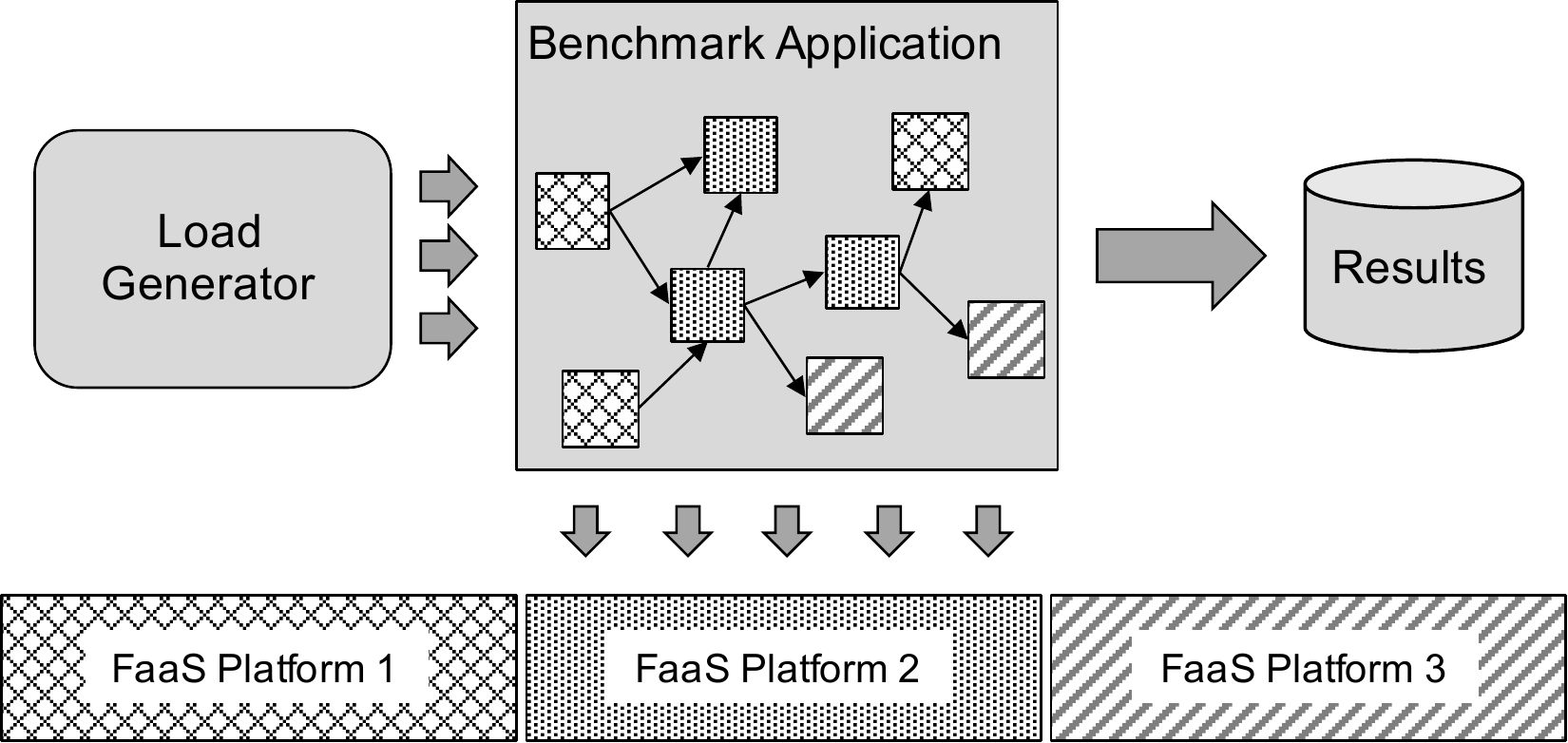}
    \caption{High-level overview of the BeFaaS architecture.}
    \label{fig:arch}
\end{figure}

%% file: fig2_overview.tex
\begin{figure}
    \centering
    \includegraphics[width=0.8\columnwidth]{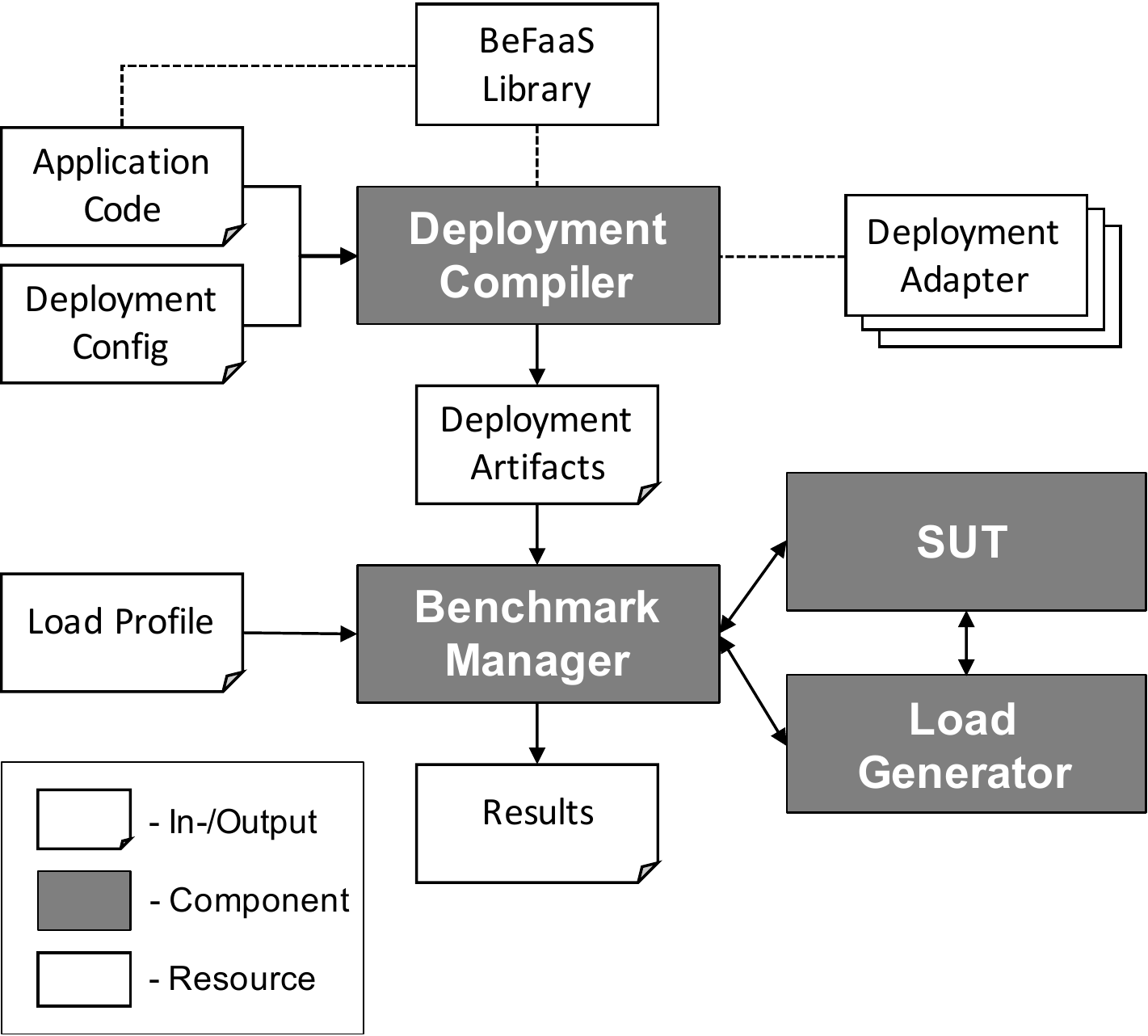}
    \caption{The Deployment Compiler transforms application code into individual deployment artifacts based on a deployment configuration. These are then deployed and benchmarked by the Load Generator. Finally, the Benchmark Manager aggregates and reports fine-grained results.}
    \label{fig:overview}
\end{figure}

%% file: 5_impl.tex
\section{Implementation}
\label{sec:impl}

Our open-source prototype implementation of BeFaaS\footnote{\url{https://github.com/Be-FaaS}} includes (i)~the BeFaaS library, (ii)~six deployment adapters, (iii)~the Deployment Compiler, (iv)~the Benchmark Manager, (v)~one realistic benchmark application, and (vi)~several load profiles for the benchmark application (see \cref{fig:overview}).

The BeFaaS library is written in JavaScript and handles calls to other functions depending on their canonical name, generates tracing IDs, and takes timestamps.
BeFaaS deployment adapters are implemented using Terraform\footnote{\url{https://www.terraform.io/}} commands.
Currently, BeFaaS thus supports three major cloud offerings (AWS Lambda, Google Cloud Functions, and Azure Functions) as well as the three open-source systems tinyFaaS~\cite{pfandzelter_tinyfaas_2020}, OpenFaaS, and OpenWhisk~\cite{baldini2017serverless} which support the deployment of functions on private infrastructure, including edge or fog nodes.
The Deployment Compiler is a shell script that uses several tools to build the deployment adapters for the respective platforms, parses and injects information from the Deployment Configuration, and generates the deployment artifacts from the application code.
The Benchmark Manager uses Terraform to create the infrastructure based on these artifacts, collect the logs, and later remove provisioned resources.
The implemented benchmark application is written in JavaScript and includes calls to external services such as a Redis\footnote{\url{https://redis.io/}} instance.
The Load Generator uses Artillery\footnote{\url{https://artillery.io/}} to call the benchmark application.
New load profiles can easily be added by specifying new Artillery load descriptions (YAML\footnote{\url{https://yaml.org/}} configuration files).

\free
\textbf{E-Commerce Application (Webshop)}

Our e-commerce benchmark implements a webshop as a FaaS application based on Google's microservice demo application\footnote{\url{https://github.com/GoogleCloudPlatform/microservices-demo}}.
Our corresponding benchmark implementation follows a typical request-response invocation style, comprises 17 functions, and uses a Redis instance as an external service to persist state (see ~\cref{fig:ecomm}).
Besides functions that provide recommendations and advertising, customers can log in, set their preferred currency, view products, fill a virtual shopping cart, check out orders, and finally observe the shipping.
Each task is implemented in a separate function (in the figure, we grouped some functions to increase legibility) and all requests arrive at a single function, the frontend, which takes the customer calls and routes them to the respective backend functions.
There are blocking synchronous calls to other functions as well as asynchronous call blocks that idle until all functions returned.

\input{fig3_ecomm.tex}

The Load Generator for the benchmark application uses Artillery running in a Docker container that can be deployed on an arbitrary instance.
It either executes a realistic default load profile that stresses all relevant aspects of the application or specific additional load profiles that emphasize stress situations, e.g., to provoke more cold starts.
The default load profile simulates four different customer workflows and constant traffic for 15~minutes.
The benchmark also includes alternative load profiles for a growth workload which linearly ramps up the load to 20 workflows per second over 15 minutes and a spike workload which suddenly increases the load from 3.5 to 20 workflows per second after five minutes, retains the high load for ten minutes, and finally continues with the lower load (3.5 workflows per second) for five minutes.

Our e-commerce benchmark is particularly well suited for comparing different cloud providers but can also be used to explore federated cloud deployments, e.g., for scenarios in which the application is running on multiple cloud platforms.

%% file: fig3_ecomm.tex
\begin{figure}
  \centering
	\includegraphics[width=\columnwidth]{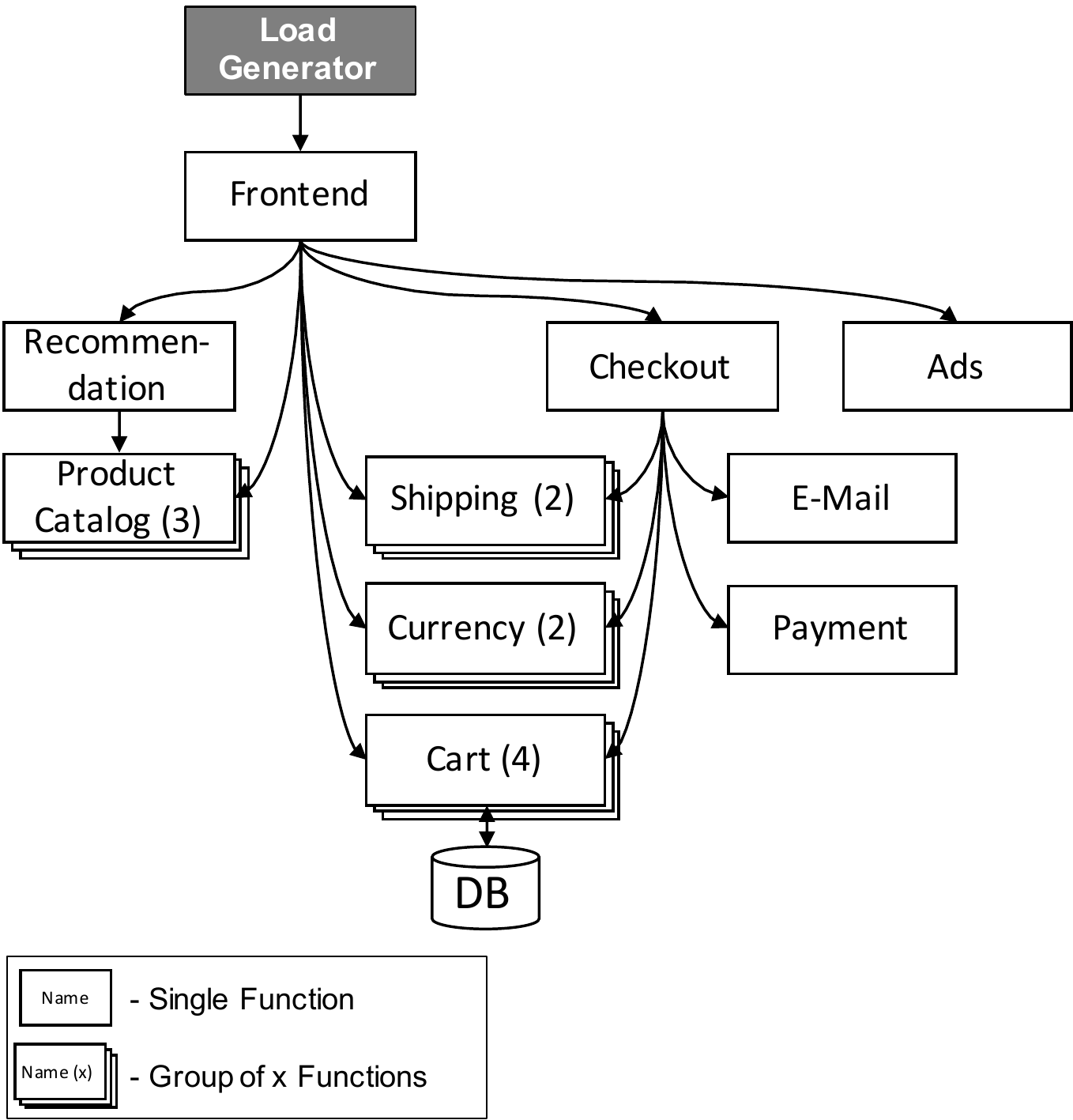}
  \caption{The e-commerce application implements a webshop in 17 functions. The frontend serves as a single entry point and an external database is used to store state.}
	\label{fig:ecomm}
\end{figure}

%% file: 6_evaluation.tex
\section{Evaluation}
\label{sec:evaluation}
We evaluate BeFaaS in two different ways. We start by presenting the results of several experiments in which we use BeFaaS to stress different FaaS platforms (\cref{subsec:experiments}). Afterwards, in \cref{subsec:disc-reqs}, we discuss to which degree BeFaaS fulfills our requirements from \cref{sec:requirements}.

\subsection{Experiments\label{subsec:experiments}}
To show how BeFaaS can be used to evaluate cloud FaaS platforms, we deploy BeFaaS in  single cloud provider setups in which all functions of the respective benchmark application are deployed on a single provider.
Here, we deploy the e-commerce benchmark on three major cloud providers (namely AWS, Azure, and GCP) and use the default load profile to compare them.

\subsubsection{Setup}
\Cref{fig:eval1} shows the basic setup of our cloud experiments:
We deploy the Load Generator on a (vastly over-provisioned) virtual machine (2~vCPUs and 4~GB~RAM) and let it execute the default load profile against the e-commerce application deployed in either \textit{eu-west-1} for AWS, \textit{westeurope} for Azure, or \textit{europe-west1} for GCP.
Moreover, the Redis database system used by the webshop also runs on an over-provisioned virtual machine (2~vCPUs and 4~GB~RAM; \textit{ta3.medium} at AWS, \textit{Standard\_B2S} in Azure, and \textit{e2-medium} at GCP) at the respective provider site.
This ensures that the database instance and Load Generator will not be a bottleneck during the experiment~\cite{book_cloud_service_benchmarking}.
During each experiment, the Load Generator executes $18,000$ workflows, which each consist of 1 to 9 requests, over a time span of 15 minutes.
Since the focus of this paper is on BeFaaS and its features and not on providing an in-depth performance analysis of different cloud providers, we decided not to repeat the experiment several times.

\input{fig5_eval1.tex}

\subsubsection{Results}
\Cref{fig:eval1result1} shows the execution duration of four selected functions which are called from the frontend function (as boxplots, boxes represent quartiles, whiskers show the minimum and maximum values without outliers beyond 1.5 times the Inter Quartile Range).
For the four functions examined in more detail, the overall picture is similar for all three providers:
As expected, simpler functions that only read or write a single value have a lower execution duration than more complex ones such as the \textit{getCart()} or \textit{checkout()} function.
In our experiment, Azure provided the fastest environment for this single run while AWS was notably slower with higher variance.

\input{fig6_functionRuntimes.tex}

In a further fine-grained analysis, we investigate the distribution of computing, network transmission, and database query latency for a synchronous and blocking section which involves two functions and database operations when an item is put into the shopping cart.
For this evaluation, we consider the (i) computation part as function runtime without the duration of outgoing network calls, (ii) network latency as the duration of outgoing calls to other function without the runtime of the called function itself, and (iii) query latency as the duration of calls to the external database. 
The detailed timestamp mechanisms of BeFaaS allow us to easily separate these times, which are shown in \cref{fig:eval1result2}.
Even though the results are based on only one benchmark run, it is noticeable that for all providers time is mostly spent on network transmission followed by the database round-trip time while the actual computing time is relatively low.
In addition, AWS showed the fastest compute and database times for the experiment run compared to the other providers Azure and GCP, while also incurring significantly more time for network transmission.

\input{fig7_computeNetworkDB.tex}

\subsection{Discussion of Requirements\label{subsec:disc-reqs}}
In \cref{sec:requirements}, we had identified six requirements for application-centric FaaS benchmarking frameworks.
We now discuss to which degree BeFaaS fulfills these requirements.

BeFaaS already comes with one standard benchmark that covers a representative FaaS application scenario, namely standard web applications, and can be easily extended by implementing more FaaS application scenarios using the BeFaaS library.
We, hence, believe that BeFaaS fulfills the requirements \textbf{R1} (\textit{Realistic Benchmark Application}) and \textbf{R2} (\textit{Extensibility for New Workloads}).

In BeFaaS, benchmark users can define arbitrarily complex deployment mappings of functions to target FaaS platforms including federated multi-cloud setups or mixed cloud/edge/fog deployments.
In fact, each function could run on a different platform. To achieve this, BeFaaS transforms the benchmark application into deployment artifacts fitted to the target platform.
Adding another target platform is also straightforward and only requires the benchmark user to implement an adapter component for the respective FaaS platform or to copy and adapt an existing adapter component.
Based on this, we argue that BeFaaS fulfills the requirements \textbf{R3} (\textit{Support for Modern Deployments}) and \textbf{R4} (\textit{Extensibility for New Platforms}).

At runtime, BeFaaS collects fine-grained measurements and traces individual requests similar to what Dapper~\cite{sigelman2010dapper} does for microservice applications.
This offers the necessary information basis for drill-down analysis.
Beyond this, BeFaaS also offers visualization capabilities for select standard measurements to further support analysis needs.
Overall, we hence conclude that BeFaaS addresses requirement \textbf{R5} (\textit{Support for Drill-down Analysis}).

Finally, we believe that BeFaaS is easy to use due to its experiment automation features and requires only very few configuration files (requirement \textbf{R6} -- \textit{Minimum Required Configuration Overhead}).
Nevertheless, this is a highly subjective matter that depends on the respective individual.
Therefore, we invite all researchers to use our proof-of-concept prototype\footnote{\url{https://github.com/Be-FaaS}} and to try it out themselves.

%% file: fig5_eval1.tex
\begin{figure}
    \centering
    \includegraphics[width=0.5\columnwidth]{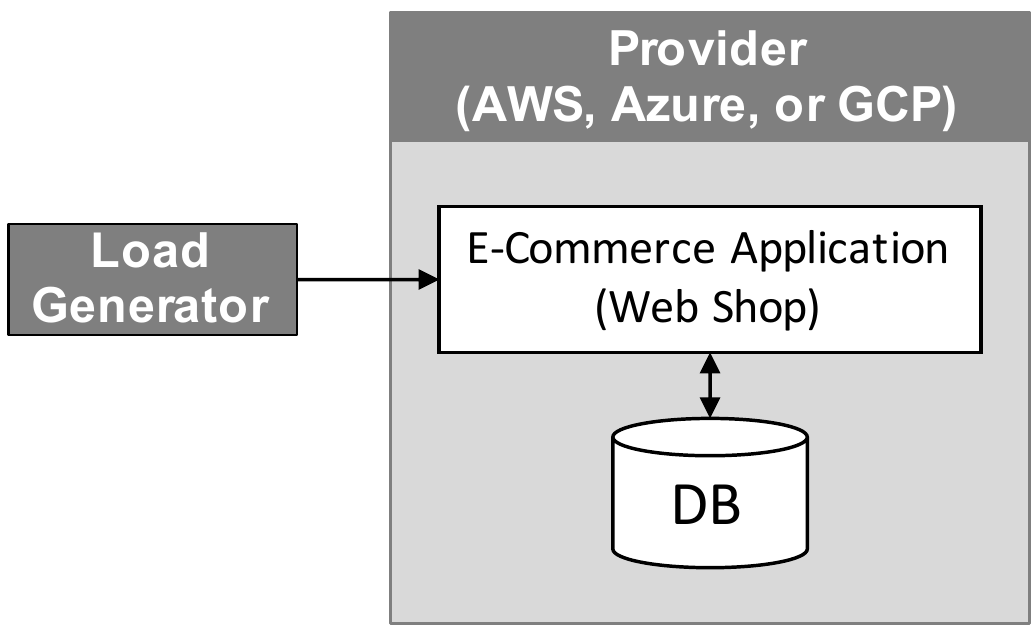}
    \caption{As part of the FaaS application, the database instance is deployed in the same region and on the same provider as the rest of the webshop.}
    \label{fig:eval1}
\end{figure}

%% file: fig6_functionRuntimes.tex
\begin{figure}
    \centering
    \includegraphics[width=\columnwidth]{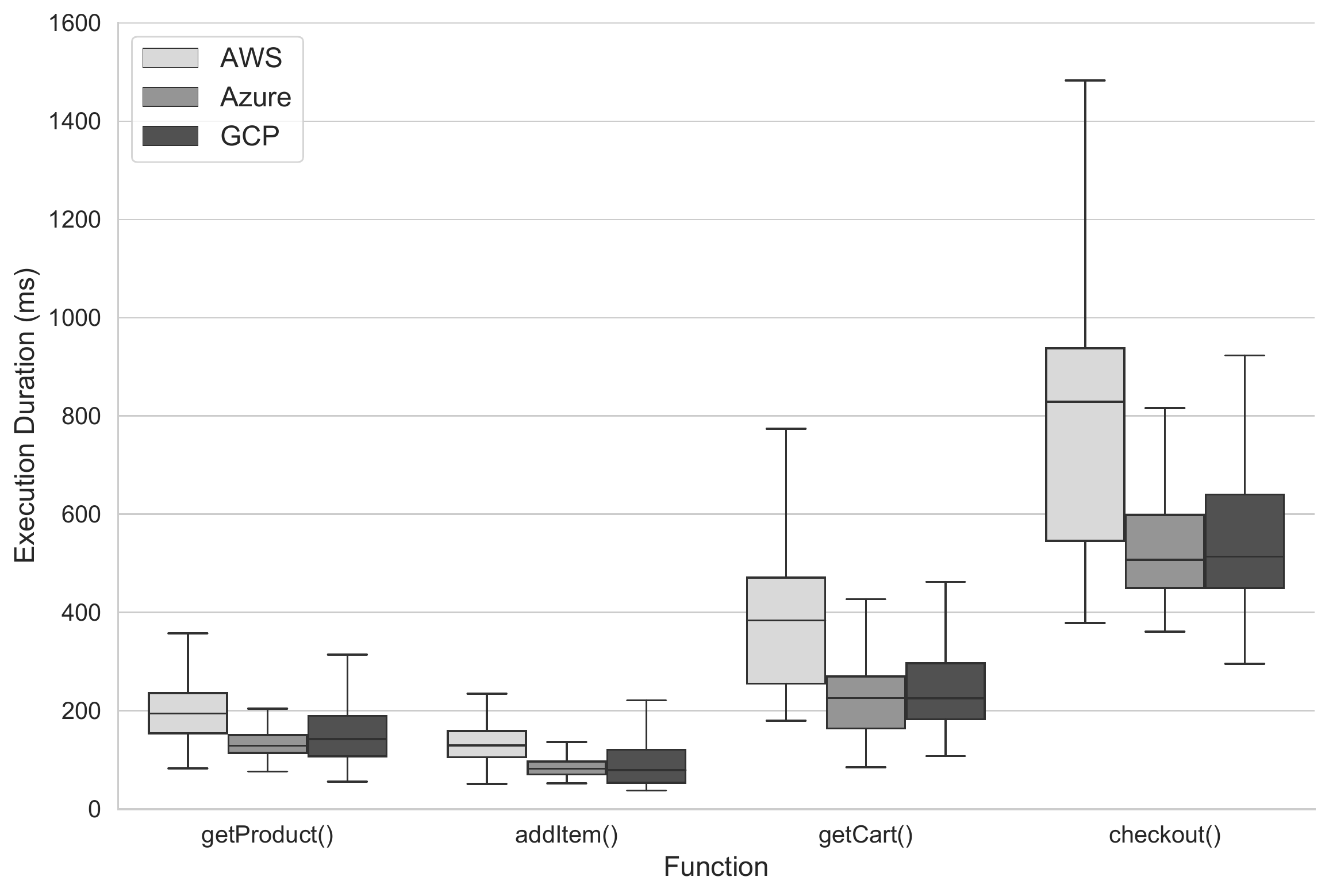}
    \caption{A detailed analysis of four functions called from the frontend shows that Azure provides the best performance and that the execution duration has the highest variance on AWS.}
    \label{fig:eval1result1}
\end{figure}

%% file: fig7_computeNetworkDB.tex
\begin{figure}
  \centering
	\includegraphics[width=\columnwidth]{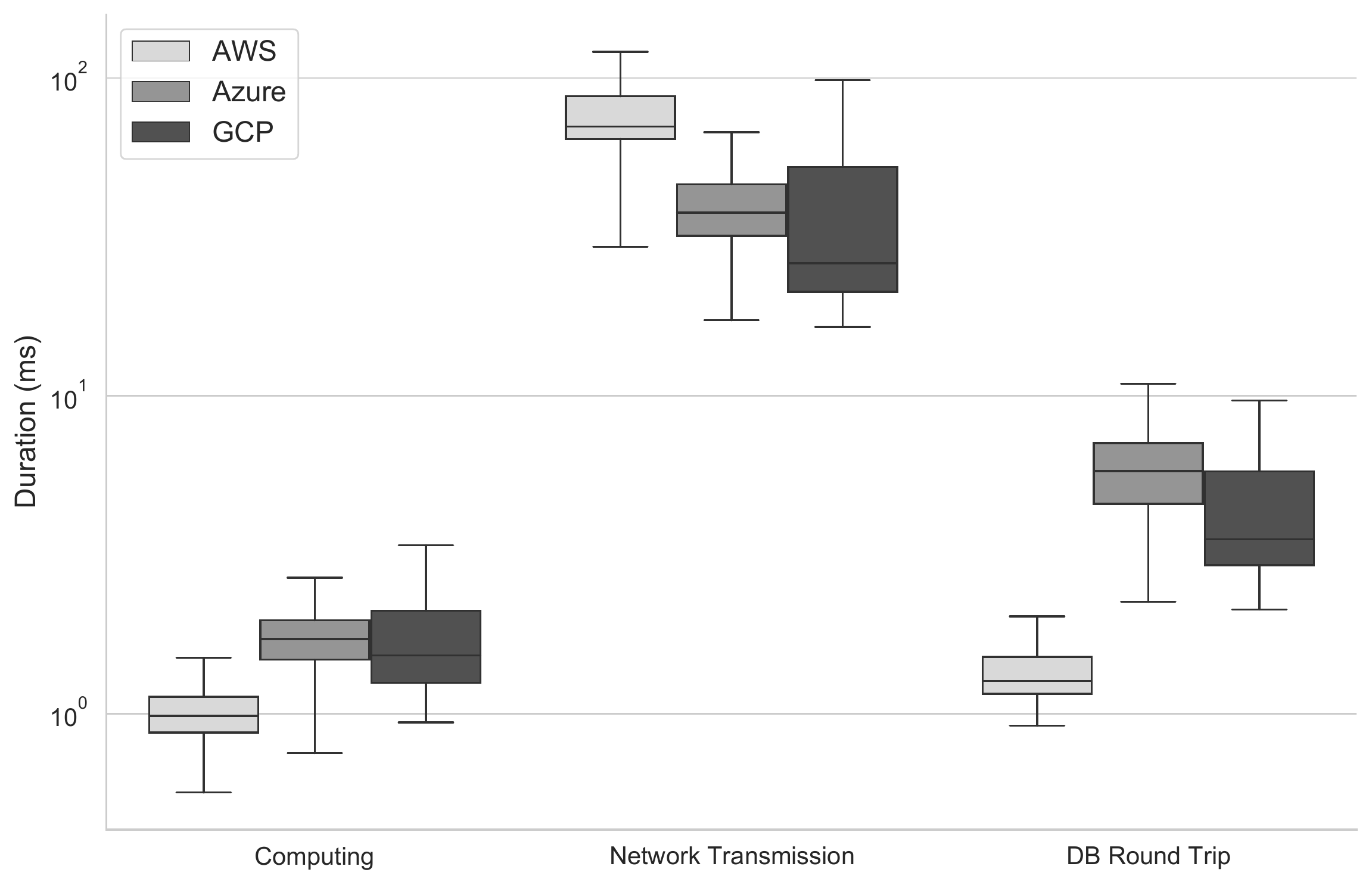}
  \caption{A drill-down analysis of a function sequence reveals that the network transmission time is the most relevant driver of execution time on all providers.}
	\label{fig:eval1result2}
\end{figure}

%% file: 7_discussion.tex
\section{Discussion}
\label{sec:discussion}
BeFaaS is a powerful, modern application-centric FaaS benchmarking framework. 
There are, however, also some points to consider and limits when using BeFaaS. 

\free
\textbf{Tracing token generates constant network overhead.}
BeFaaS supports a detailed tracing of requests by injecting a small token in each call. 
On the one hand, this supports the clear mapping of different calls to function chains, yet on the other hand, it also causes an additional network overhead.
This token, however, is mostly constant in size (depending on the length of the respective function name), so the overhead can be easily determined and considered in results analysis.
Furthermore, this will only matter if the goal of the benchmark is to find the optimal deployment for an existing application which is then instrumented to be used as a BeFaaS benchmark.

\free
\textbf{No detailed measurements for external services.}
Currently, BeFaaS handles external services and components as a black-box and only measures end-to-end latency of such service calls.
In future work, however, we plan to implement a small BeFaaS sidecar proxy that can be deployed on external service instances to forward calls to the respective service and to inject the BeFaaS tracing token there as well. 

\free
\textbf{External services can affect the comparability and fairness.}
The included benchmark uses an external database system to persist state but further benchmarks and use cases may also require external services such as pub/sub message brokers or web APIs.
Although the modular design of BeFaaS supports this, there are also some pitfalls in terms of fairness and comparability:
In our experiments, we placed the database instance in the same region and deployed it at the same provider to minimize latency between functions and database.
A function calling the external service and awaiting the response will not idle for a long time and the execution environment at the provider side will soon be available again for the next request. 
On the other hand, a function calling an external service in another region with larger latency will block the environment and (may) cause a cold start for the next incoming request. 
Thus, when using external services, these should be located and deployed with similar latency for all alternatives.
Moreover, as cloud environments are virtually infinitely scalable, it has to be assured that the external service does not become a performance bottleneck during the experiment.
Otherwise, the benchmark would benchmark the compute resources of the external service instead of the FaaS environment.

\free
\textbf{Provider-specific features can affect portability.}
Competing FaaS vendors are constantly developing new and exclusive features that simplify development and deployment for customers. 
These features, however, can also affect the portability of the BeFaaS framework if a (future) benchmark uses exclusive features that are not available at all vendors.
Thus, we strongly recommend not to use exclusive features of individual providers when developing new BeFaaS benchmarks.
BeFaaS can, however, help to determine the impact of new features within a provider or across multiple providers by adjusting and configuring the respective deployment adapter.  

\free
\textbf{Clock synchronization is required for some drill-down analysis tasks.}
The drill-down analysis features of BeFaaS require approximately synchronized clocks.
Although this will usually be provided by the provider with sufficient accuracy, a user should assert this before running experiments as this will affect the reliability of tracing insights.
Nevertheless, such detailed insights may often not be needed and the tracing of BeFaaS also offers something to counteract this:
If the call follows a request-response pattern, BeFaaS measures the total round trip time at the calling function and knows the computing duration at the called function. 
Thus, it is possible to approximate the network transmission latency under the assumptions that both directions took comparably long.
For event-based calls that do not return a message to the sender, however, this is not possible. 
In our experience, though, this is not a problem in the cloud and for self-hosted FaaS platforms, where the user has direct control over clock synchronization.

%% file: 8_conclusion.tex
\section{Conclusion}
\label{sec:conclusion}
FaaS platforms are a popular cloud compute paradigm and have also been proposed for edge environments.
For comparing and choosing different FaaS platforms in terms of performance, developers usually rely on benchmarking.
Existing FaaS benchmarks, however, tend to fall into the microbenchmark category -- an application-centric FaaS benchmarking framework is still missing.

In this paper, we presented BeFaaS, an extensible framework for executing application-centric benchmarks against FaaS platforms which comes with a realistic e-commerce benchmark.
BeFaaS is also the first benchmarking framework with out-of-the-box support for federated cloud setups which allows us to also evaluate complex configurations in which an application is distributed over multiple FaaS platforms running on a mixture of cloud, edge, and fog nodes.
We plan to use this feature of BeFaaS in future work to evaluate the feasibility of such infrastructure.
Beyond this, BeFaaS is focused on ease-of-use through automation and collects fine-grained measurements which can be used for a detailed post-experiment drill-down analysis, e.g., to identify cold starts or other request-level effects; it can easily be extended with additional benchmarks or adapters for further FaaS platforms.

With BeFaaS, we provide developers with the necessary tool to explore, compare, and analyze FaaS platforms for their suitability for application scenarios.
We also offer researchers the ability to study the performance effects of different FaaS deployment options across cloud, edge, and fog through experiments.